\documentclass{article}

\newcommand{\p}[1]{(\ref{#1})}

\newcommand{\bX}{{\overline X}}

\newcommand{\bD}{{\overline D}}

\newcommand{\bW}{{\overline W}}

\newcommand{\bxi}{{\bar\xi}}

\newcommand{\eps}{\varepsilon}

\newcommand{\be}{\begin{equation}}
\newcommand{\ee}{\end{equation}}
\newcommand{\bea}{\begin{eqnarray}}
\newcommand{\eea}{\end{eqnarray}}

\newcommand{\ba}{\begin{array}} \newcommand{\ea}{\end{array}}

\def\im{{\rm i}}

\newcommand{\nn}{\nonumber}

\usepackage{amscd,amsmath,amssymb}

\begin{document}
\thispagestyle{empty}
\vspace{2cm}
\begin{flushright}
\end{flushright}\vspace{2cm}
\begin{center}
{\Large\bf Ferrara-Porrati-Sagnotti approach and the one-dimensional\\
\vspace{0.5cm}
 supersymmetric model with PBGS}
\end{center}
\vspace{1cm}

\begin{center}
{\large\bf S.~Bellucci${}^a$, S.~Krivonos${}^{b}$,
A.~Sutulin${}^{a,b}$}
\end{center}

\begin{center}
${}^a$ {\it
INFN-Laboratori Nazionali di Frascati,
Via E. Fermi 40, 00044 Frascati, Italy} \vspace{0.2cm}

${}^b$ {\it
Bogoliubov  Laboratory of Theoretical Physics, JINR,
141980 Dubna, Russia} \vspace{0.2cm}

\end{center}
\vspace{2cm}

\begin{abstract}\noindent
We apply Ferrara-Porrati-Sagnotti approach to the case of one-dimensional supersymmetric model with
$N=2$ supersymmetry spontaneously broken to the $N=1$ one.
We explicitly demonstrate that only one superfield can be treated as the Goldstone one, while the second one has the meaning of the matter superfield.
The general action for such a system is constructed and also two relevant particular cases are considered.
\end{abstract}

\newpage
\setcounter{page}{1}
\setcounter{equation}{0}

\section{Introduction}
In \cite{BG} {\it J.~Bagger and A.~Galperin} ({\it BG}) proposed the approach to construct the $N=2, D=4$ supersymmetric Born-Infeld theory.
Combining two $N=1$ supermultiplets, namely, $N=1$ {\it vector} and  $N=1$ {\it chiral} ones, they realized the
transformations of additional $N=1$ supersymmetry with  parameters $\eta_{\alpha}, \bar\eta_{\dot\alpha}$ as
\be\label{i1}
\delta \left( W\right)_\alpha = \left( 1-\frac{1}{4} \bD{}^2 \bX \right)\eta_\alpha -\im \partial_{\alpha\dot\alpha} X \bar\eta{}^{\dot\alpha}. \quad
\delta X = -2 \left( W \right)^\alpha \eta_\alpha,
\ee
Due to the presence of the constant term in this transformation law, this additional supersymmetry is spontaneously broken.
Keeping in mind, that from the vector superfield $W_\alpha$ Bianchi identity, for the $N=1$ vector supermultiplet it follows that
\be\label{x2}
D^2 W_\alpha \sim \partial_{\alpha\dot\alpha} \bW^{\dot\alpha},
\ee
then the trivial action
\be\label{i2}
S= \int d^4x d^2\; \theta X
\ee
becomes invariant under transformations \p{i1}. However, the action \p{i2} acquires a real meaning only after introducing the additional constraint
(which is invariant under \p{i1})
\be\label{i3}
W\cdot W +X \left( 1 -\frac{1}{4} \bD{}^2 \bX \right) =0.
\ee
Once again, due to the presence of the constant in \p{i3}, this constraint can be explicitly solved by expressing $X$ in terms of the superfields $W_\alpha, \bW_{\dot\alpha}$. The resulting action is just the $N=2$ Born Infeld one.

In the recent paper \cite{FPS} {\it S.~Ferrara, M.~Porrati and A.~Sagnotti} ({\it FPS}) proposed the generalization
of the {\it BG} construction to the cases
of several $N=1$ vector supermultiplets.
The {\it FPS} approach includes {\it two} basic ingredients.

Firstly, the nonlinear constraint \p{i3} is generalized to be
\be\label{i4}
d_{abc} \left( W_b \cdot W_c +X_b \left( m_c -\frac{1}{4} \bD{}^2 \bX_c \right)\right) =0,
\ee
where $W_{a \alpha}$ and  $X_a$ are $n$-copies of $N=1$ vector and chiral multiplets respectively,
and the totally symmetric tensor $d_{abc}$ and vector $m_a$ are a set of constants.

The realization of hidden $N=1$ supersymmetry is similar to the {\it BG} case and reads
\be\label{i5}
\delta  W_{a \alpha} = \left( m_a -\frac{1}{4} \bD{}^2 \bX_a \right)\eta_\alpha -\im \partial_{\alpha\dot\alpha} X_a \bar\eta{}^{\dot\alpha}, \quad
\delta X_a = -2 W_a^\alpha \eta_\alpha.
\ee
The invariance of \p{i4} with respect to \p{i5} involves the new {\it additional} constraint
\be\label{i6}
d_{abc} W_{b \alpha}  X_c =0,
\ee
which was automatically satisfied in the BG case. The constraints \p{i4}, \p{i6} can be solved to express the bosonic $N=1$ chiral superfields $X_a$
through the $N=1$ vector supermultiplets $W_{a \alpha}$.

The next nontrivial step of the {\it FPS} approach yields the structure of the corresponding action, which proved to be more involved
\be\label{i7}
S= \int d^4x d^2 \theta \;\left[  e^a X_a+ C_{ab} \left( W_a \cdot W_b +X_a \left( m_b -\frac{1}{4} \bD{}^2 \bX_b \right)\right)+ c.c.\right],
\ee
where $e^a = const, C_{ab}=C_{ba}=const.$
In fact, the additional term with $C_{ab}$ is invariant with respect to $S$-transformations \p{i5}. Such a term does not exist
in the case of one supermultiplet, but it proved to be {\it essential} in the cases of several supermultiplets.
The action \p{i7} with $X_a$ being the solution of the constraints \p{i4}, \p{i6} is treated as
the many-field extension of $N=2$ Born-Infeld theory.

In \cite{FPS} the detailed analysis of the $n=2$
case was presented, which can be divided into two subcases:
\begin{itemize}
\item $d_{111}=1, d_{112}=-1 \quad I_4=0$,
\item $d_{111}=1, d_{122}=-1 \quad I_4>0$,
\end{itemize}
where $I_4$ is a quartic invariant. Then, the analysis of the next $n=3$ case has been given in \cite{FPS1}.

\noindent
The above, quite short sketch of the {\it FPS} approach is enough to raise {\it two} interesting questions:
\begin{itemize}
\item whether the supermultiplets with {\it different} constants $d_{abc}$ are really {\it different?}
\item whether the action \p{i7} is {\it unique} as it happened in the $n=1$ case?
\end{itemize}

The reasons for raising the first question are the following.

\vspace{0.3cm}
After resolving the constraints \p{i4}, \p{i6},
we will have in the theory $n$ copies of $N=1$ vector multiplets $W_{a \alpha}$ with
highly nonlinear transformation properties
under hidden $N=1$ supersymmetry \p{i5}. The presence of the constants $m_a$ in \p{i5} at first sight means that
all of them have Goldstone nature. But it cannot be so, because the spontaneous breaking of $N=2$ to $N=1$
supersymmetries is accompanied by only one Goldstone superfield \cite{NLR1,NLR2}.
Therefore, it should exist a special basis in which
only one superfield is a Goldstone one, while the remaining superfields have to be matter
superfields with trivial transformation properties with respect to the broken $S$-supersymmetry.
Due to close relations between linear and nonlinear realizations of partially broken supersymmetries \cite{ikap1},
the same property has to be visible within the linear realizations method, the {\it FPS} approach is dealing with.
Thus, the claim that different choices of the constants $d_{abc}$
in the basic constraints \p{i4}, \p{i6} must be equivalent, should be carefully analyzed.

The second question about the uniqueness of the action we are raising, is also related
with the existence of the special basis in which we have one Goldstone superfield and an arbitrary number of matter superfields.
In such a basis the general action will have a functional freedom. Thus, it has to be quite interesting to analyze this general action and understand
the reasons that just select the {\it FPS} actions.

Unfortunately, the analysis of $N=2$ extended Born-Infeld theory is quite involved, and the ideological part is hidden
behind complicated calculations. Fortunately enough, the {\it FPS} approach, as we demonstrated in \cite{comments}, is not limited by application to the Born-Infeld theory only, and the questions we raised above may be answered through the analysis of a much simpler system, which the {\it FPS} approach can be applied to.

In our paper \cite{BKS-FPS} we explored the generalization of the {\it FPS} construction for
the simplest model of super-particle which realized the  partial breaking of $N=2$ to $N=1$ supersymmetry in one dimension.
We explicitly established the transformations that relate two different $n=2$ cases with the special basis discussed above, as well as
the relations between them.  We also constructed the general action for such a system containing
two arbitrary functions and provided the explicit form of these functions reducing the general action to {\it FPS}-like ones.

In this contribution we review our results.

\setcounter{equation}{0}
\section{Partial breaking in one dimension and the superparticle action}
We begin  with a simple example of a superparticle model describing $N=2\; \rightarrow\; N=1$ partial
breaking of global supersymmetry in $d=1$, which was considered in paper \cite{DIK}
The anticommutation relation of $N=1, d=1$ Poincar\'e superalgebra reads
\be\label{alg}
\big \{Q, Q \big \} = 2P\,,
\ee
and the coordinates $(t, \theta)$ of $N=1$ superspace satisfy the following rules of complex conjugation:
$t^{\dagger} = -t, \theta^{\dagger} = \theta$.
Now we define the bosonic and fermionic superfields $v(t, \theta)$ and $\psi(t, \theta)$ related as
\be\label{SF}
\psi = \frac{1}{2}\, Dv, \quad (v^{\dagger} = -v,\;  \psi^{\dagger} = \psi),
\ee
where the spinor derivative $D$ satisfies the relation
\be\label{spinor-der}
D = \frac{\partial}{\partial \theta} + \theta \partial_t, \quad
\big \{D, D \big \} = 2 \partial_t\,.
\ee

\subsection{Linear realization: one-particle case}
In full analogy with {\it BG} approach \cite{BG}, we introduce two spinor superfields
$\psi(t, \theta)$ and $\nu(t, \theta)$, assuming that an additional spontaneously broken $N=1$ supersymmetry is realized on them as
\be\label{susy-S}
\delta \psi = \eps (1 - D \nu)\,, \quad \delta \nu = \eps D \psi.
\ee
The presence of the constant shift in the first relation suggests the interpretation
of $\psi(t, \theta)$ as the Goldstone fermion accompanying  the $N=2 \rightarrow N=1$ breaking.
The superfield $\nu(t, \theta)$, due to its transformation properties under the broken $N=1$ supersymmetry,
may be chosen as a {\it Lagrangian density,} since the integral over the superspace
\be\label{action1}
S = \int dt d\theta\, \nu
\ee
is invariant with respect to both broken and unbroken supersymmetries.

To be meaningful, the action \p{action1} should be accompanied by
an additional constraint, which is invariant under transformations \p{susy-S}.
One may easily check, that the corresponding invariant constraint looks quite similar to \p{i3}
\be\label{constr1a}
\psi D \psi - \nu \left( 1 - D \nu\right) =0
\ee
with evident solution given by
\be\label{sol-constr1}
\nu = \frac{2 \psi D \psi}{1+\sqrt{1-4 (D\psi)^2}}\,.
\ee
It is easy to find that from \p{constr1a} it follows that
\be\label{constr11}
\psi \nu =0.
\ee
The bosonic limit of the action \p{action1} with \p{sol-constr1} corresponds to the action of a particle in $d=1$
\be\label{comp-action1}
S_{bos} = \frac{1}{2} \int dt \left( 1 - \sqrt{1- (\dot v)^2}\right).
\ee
Thus, this simplest system demonstrate a close analogy with the {\it BG} construction of the supersymmetric Born-Infeld action.

\subsection{Linear realization: the multiparticle case}
The above considered case can be easily extended to that of an arbitrary number of $N=1$ superfields strictly following the {\it FPS}
approach. Doing so, we firstly introduce a set of $n$ superfields $\psi_a(t, \theta)$ and $\nu_a(t, \theta)$
transforming under implicit $N=1$ supersymmetry as
\be\label{x1}
\delta \psi_a = \eps (m_a - D \nu_a)\,, \quad \delta \nu_a = \eps D \psi_a,
\ee
where $m_a$ are arbitrary constants.
Now, again in full analogy with {\it FPS} approach, we impose the following generalization of the constraint \p{constr1a}
\be\label{FPS-constr}
d_{abc} \big(  \psi_b D \psi_c - \nu_b \left (m_c- D \nu_c \right )\big) = 0.
\ee
The invariance of this constraint under the broken $S$-supersymmetry leads to an additional restriction
\be
\partial_t \left(d_{abc} \psi_b \nu_c \right) = 0,
\ee
which can be reformulated as
\be\label{add-constr}
d_{abc} \psi_b \nu_c = 0,
\ee
assuming that the constant of integration is equal to zero.
Finally,  the {\it FPS}-like generalization of the action reads
\be\label{action}
S = \int dt d\theta\,\Big[  e^a \nu_a+ C_{ab}  \big(  \psi_a D \psi_b - \nu_a \left (m_b- D \nu_b \right )\big)\Big],
\ee
where $e^a$ and $C_{ab}=C_{ba}$ are arbitrary real constants.

After repeating the basic steps of the {\it FPS} approach, let us  consider the system of two superparticles in the details.

\setcounter{equation}{0}
\section{Standard basis for $n=2$ cases}
In the standard approach of the Partial Breaking of Global Supersymmety, the two-particle model should contain only
one Goldstone superfield (with the Goldstone fermion among their components).
The rest of the fields must be matter ones, i.e. they should not transform
with respect to the broken S-supersymmetry.
Now we demonstrate how such a splitting works in a two-particle system
for special choices of the components of tensor $d_{abc}$. We show that there is a special basis in which the systems in question,
corresponding to a different choice of $d_{abc}$, can be described in a unified way.
\subsection{The case with $d_{111}=d_{222}=1$}
With such a choice of the symmetric tensor $d_{abc}$, the basic constraints \p{FPS-constr}, \p{add-constr} have a splitting form
\bea\label{constr1}
&&
\psi_1 D\psi_1 -\nu_1\left(m_1 - D \nu_1\right) =0, \quad   \psi_1 \nu_1 =0, \\
&&
\psi_2 D\psi_2 -\nu_2\left(m_2 - D \nu_2\right) =0,  \quad  \psi_2 \nu_2 =0. \nn
\eea
By rescaling of the variables $\psi_{a}$ and $\nu_{a}$ in \p{constr1}, one may always choose
\be\label{m1}
m_1=m_2=1,
\ee
then the transformations of fermions $\psi_{a}$ and $\nu_{a}$ in \p{x1} take the form
\bea\label{tr1}
&&
\delta \psi_1 = \epsilon \left(1 - D \nu_1\right), \quad  \delta\nu_1 =\epsilon D\psi_1, \\
&&
\delta \psi_2 = \epsilon \left(1 - D \nu_2\right), \quad  \delta\nu_2 =\epsilon D\psi_2. \nn
\eea
Similarly to the one particle case \p{sol-constr1}, the constraints \p{constr1} can be solved as
\be\label{cc2}
\nu_1 = \frac{2 \psi_1 D \psi_1}{1+\sqrt{1-4 (D\psi_1)^2}}\,, \qquad \nu_2 = \frac{2 \psi_2 D \psi_2}{1+\sqrt{1-4 (D\psi_2)^2}}\,.
\ee
The {\it FPS}-like action \p{i7}, invariant with respect to both $S$- and $Q$-supersymmetries, for this case reads
\be\label{action-tilde1}
S = \int dt d\theta \, \Big\{ e_1\,\nu_1 +e_2 \nu_2\, +C_{12}\big( \psi_1 D \psi_2 - \nu_1 (1 - D \nu_2)+\psi_2 D \psi_1 - \nu_2 (1 - D \nu_1)\big) \Big\}.
\ee
Now we introduce two Goldstone spinor superfields $\xi_{a}$ that transform as
\be\label{xi1}
\delta \xi_1 = \epsilon + \epsilon\, \xi_1 \, \partial_t\, \xi_1, \qquad \delta \xi_2 = \epsilon + \epsilon\, \xi_2 \, \partial_t\, \xi_2.
\ee
It is easy to check that the {\it tilded fields} $\tilde\psi_{a}$ and $\tilde\nu_{a}$
\be\label{tpsi1}
\tilde\psi_a = \psi_a -\xi_a \left(1- D\nu_a\right),\quad \tilde\nu_a =\nu_a - \xi_a D\psi_a,
\ee
transform under \p{tr1}, \p{xi1} as
\be\label{tr2}
\delta \tilde\psi_a =\epsilon\, \xi_a\, \partial_t\, \tilde\psi_a, \qquad \delta \tilde\nu_a = \epsilon\, \xi_a \,\partial_t\, \tilde\nu_a.
\ee
Thus, it is a covariant operation to put the superfields \p{tpsi1} equal to zero, which is equivalent to the relations
\be\label{constr111}
\psi_a-\xi_a\left(1-D \nu_a\right)=0, \qquad  \nu_a-\xi_a D \psi_a=0.
\ee
Moreover, the constraints $\tilde\psi_a=\tilde\nu_a=0$ contain some
additional information: they express the superfields of the linear realization
$\psi_a$ and $\nu_a$ in terms of the Goldstone superfields $\xi_a$
\be\label{rel1}
\nu_a = \frac{\xi_a D \xi_a}{1+D\xi_a D \xi_a},\quad \psi_a =\frac{\xi_a}{1+D\xi_a D\xi_a} \qquad \qquad \mbox{(no summation over {\it a})}.
\ee
Until now we have {\it two} Goldstone superfields $\xi_{a}(t,\theta)$, while we are expecting to
have {\it only one} essential Goldstone fermionic superfield and {\it one matter} superfield.
This may be achieved by passing to the {\it new} superfields $\eta(t,\theta)$ and $\lambda(t,\theta)$
\be\label{eta1}
\eta= \frac{1}{2}\left(\xi_1+\xi_2\right) +\frac{1}{4} \xi_1 \xi_2 \left( \dot\xi{}_1 -\dot \xi{}_2\right), \quad
\lambda = \frac{1}{2}\left(\xi_1-\xi_2\right) +\frac{1}{4} \xi_1 \xi_2 \left( \dot \xi{}_1 +\dot \xi{}_2\right),
\ee
which, in virtue of \p{xi1}, transform as expected\footnote{This is the form-variation of the fields under implicit $N=1$ supersymmetry:
$\delta {\cal A}= {\cal A}'(t,\theta)-{\cal A}(t,\theta)$.}
\be\label{eta2}
\delta \eta = \epsilon+\epsilon \eta \partial_t \eta, \quad
\delta \lambda = \epsilon \eta \partial_t \lambda.
\ee
From \p{eta2} it follows that $\eta(t,\theta)$ is the fermionic Goldstone superfield accompanying the  spontaneous
breaking of $N=2$ supersymmetry, while $\lambda(t,\theta)$ is a matter fermionic superfield.
Finally, the superfields $\xi_a(t,\theta)$ as well as $\psi_a(t,\theta)$ and $\nu_a(t,\theta)$ can be expressed in
terms of $\eta(t,\theta)$ and $\lambda(t,\theta)$:
\be\label{xi11}
\xi_1 = \eta + \lambda +\eta \lambda \big( \dot \eta + \dot \lambda \big), \quad
\xi_2 = \eta - \lambda -\eta \lambda \big( \dot \eta - \dot \lambda \big),
\ee
\bea\label{fin1}
&&
\psi_1 =\frac{\eta+\lambda}{1+\left( D\eta+D\lambda\right)^2}+ \eta\lambda \big(\dot \eta+\dot \lambda\big)
\frac{1- \left( D\eta+D\lambda\right)^2}{\big[ 1+\left( D\eta+D\lambda\right)^2\big]^2}\,,\quad
\psi_2 = \left.\psi_1\right|_{\lambda \rightarrow -\lambda}, \\
&&
\nu_1 =\left( D\eta+D\lambda\right)\bigg( \frac{\eta+\lambda}{1+\left( D\eta+D\lambda\right)^2}+
\frac{2\eta\lambda \big(\dot \eta+\dot \lambda\big)}{\big[ 1+\left( D\eta+D\lambda\right)^2\big]^2}\bigg)\,, \quad
\nu_2 = \left.\nu_1\right|_{\lambda \rightarrow -\lambda}. \nn
\eea
Thus, the system of constraints \p{constr1} represents a {\it non-standard} description of the essential
Goldstone superfield $\eta(t,\theta)$ and a matter fermionic superfield $\lambda(t,\theta)$.

\subsection{The case with $d_{111}=1, d_{122}=-1$}
For these  values of the constants the constraints \p{FPS-constr} read
\bea\label{d-FPS-constr}
&&
\psi_1 D \psi_1 - \psi_2 D \psi_2 - \nu_1 \left(m_1 - D \nu_1\right) + \nu_2 \left(m_2 - D \nu_2 \right) = 0, \\
&&
\psi_1 D \psi_2 + \psi_2 D \psi_1 - \nu_1 \left(m_2 - D \nu_2\right) - \nu_2 \left(m_1 - D \nu_1 \right) = 0, \nn
\eea
while those in \p{add-constr} take form
\be\label{d-add-constr}
\psi_1 \nu_1 - \psi_2 \nu_2 = 0, \qquad \psi_1 \nu_2 + \psi_2 \nu_1 = 0.
\ee
Now we introduce the complex superfields $\psi, \nu$ and complex parameter $m$ as
\bea\label{new-basis0}
&&
\psi = \psi_1 + \im \psi_2\,, \qquad \nu = \nu_1 + \im \nu_2\,, \qquad m = m_1 + \im m_2\,,\\
&&
\bar\psi = \psi_1 - \im \psi_2\,, \qquad \bar\nu = \nu_1 - \im \nu_2\,, \qquad \bar m = m_1 - \im m_2\,, \nn
\eea
and then rewrite eqyations \p{d-FPS-constr} and \p{d-add-constr} in the splitting form
\bea\label{new-basis}
&&
\psi D \psi - \nu \left( m - D \nu \right) = 0\,, \quad
\psi \, \nu = 0\,, \\
&&
\bar \psi D \bar\psi - \bar\nu \left( \bar m - D \bar\nu \right) = 0\,, \quad
\bar\psi \, \bar\nu = 0\,. \nn
\eea
The solution to the equations \p{new-basis} read
\be\label{sol11}
\nu = \frac{2 \psi D \psi}{m+ \sqrt{m^2 -4 (D \psi)^2}}\,, \quad \bar\nu = \frac{2 \bar\psi D \bar\psi}{\bar m+ \sqrt{\bar m^2 -4 (D \bar \psi)^2}}\,.
\ee
In terms of these variables the invariant {\it FPS} action \p{i7} acquires a form
\be\label{action-tilde}
S = \int dt d\theta\, \Big\{ e\,\nu +e^* \bar\nu\, +C_{12}\big( \psi D \bar\psi - \nu (\bar m - D \bar\nu)+\bar\psi D \psi - \bar\nu (m - D \nu)\big)\Big\}\,,
\ee
where $e=\frac{1}{2}\,(e_1- \im e_2)$.
Let us note, that one may always choose the constants $m_1, m_2$ as \cite{comments}
\be\label{m}
m_1=1, \; m_2 = 0, \qquad \Rightarrow \qquad m=\bar m=1.
\ee
With this choice of the parameter $m$, the complex superfields $\psi$ and $\nu$ transform with respect to the broken $S$-supersymmetry as
\bea\label{tr22}
&&
\delta \psi = \epsilon \left(1 - D \nu\right), \qquad  \delta\nu =\epsilon D\psi, \\
&&
\delta \bar\psi = \epsilon \left(1 - D \bar\nu \right), \qquad  \delta\bar\nu =\epsilon D\bar\psi, \nn
\eea
and obey the constraints
\bea\label{new-basis000}
&&
\psi D \psi - \nu \left( 1 - D \nu \right) = 0\,, \qquad
\psi \, \nu = 0\,, \\
&&
\bar\psi D \bar\psi - \bar\nu \left( 1 - D \bar\nu \right) = 0\,, \qquad \bar\psi \, \bar\nu = 0\,. \nn
\eea
So, we have a full analogy with the previously considered  splitting case.

\noindent
Let us introduce two Goldstone spinor superfields $\xi(t,\theta)$ and $\bar\xi(t,\theta)$ which transform as
\be\label{xi22}
\delta \xi = \epsilon + \epsilon\, \xi \, \partial_t\, \xi, \qquad \delta \bar\xi = \epsilon + \epsilon\, \bar\xi \, \partial_t\, \bar\xi.
\ee
Then the {\it tilded fields} $\tilde\psi, \bar{\tilde\psi}$ and $\tilde\nu,\bar{\tilde\nu}$, which are defined by the following expressions
\be\label{tpsi22}
\tilde\psi = \psi -\xi \left(1- D\nu\right),\quad \tilde\nu =\nu - \xi D\psi, \quad
\bar{\tilde\psi} = \bar\psi -\bar\xi \left(1- D\bar\nu\right), \quad \bar{\tilde\nu} =\bar\nu - \bar\xi D\bar\psi,
\ee
transform under \p{tr22}, \p{xi22} as
\be\label{tr23}
\delta \tilde\psi =\epsilon\, \xi\, \partial_t\, \tilde\psi, \qquad \delta \tilde\nu = \epsilon\, \xi \,\partial_t\, \tilde\nu, \qquad
\delta \bar{\tilde\psi} =\epsilon\, \bar\xi\, \partial_t\, \bar{\tilde\psi}, \qquad \delta \bar{\tilde\nu} = \epsilon\, \bar\xi \,\partial_t\, \bar{\tilde\nu}.
\ee
Thus, one may again  to put these superfields equal to zero, that means
\be\label{constr22}
\psi-\xi\left(1-D \nu\right)=0,\;\; \bar\psi-\bar\xi\left(1-D \bar\nu\right)=0, \;\; \nu-\xi D \psi=0, \;\;
\bar\nu-\bar\xi D \bar\psi=0.
\ee
It is obvious that the constraints \p{new-basis000} follow from \p{constr22}.
The constraints \p{constr22} can be solved as
\be\label{rel22}
\nu = \frac{\xi D \xi}{1+D\xi D \xi}\,,\quad \psi =\frac{\xi}{1+D\xi D\xi}\,, \quad
\bar\nu = \frac{\bar\xi D \bar\xi}{1+D\bar\xi D \bar\xi}\,,\quad \bar\psi =\frac{\bar\xi}{1+D\bar\xi D\bar\xi}\,.
\ee
In full analogy with the previously considered case,
we introduce the superfields $\eta(t, \theta)$ and $\lambda(t, \theta)$ by
\be\label{eta21}
\eta= \frac{1}{2}\big(\xi+\bar\xi\big) +\frac{1}{4}\, \xi \bar\xi \big( \dot \xi -\dot{\bar\xi}\big), \quad
\lambda = \frac{\im}{2}\, \big(\xi-\bar\xi\big) +\frac{\im}{4}\, \xi \bar\xi \big( \dot \xi + \dot{\bar\xi}\big),
\ee
and find that they have the same transformation properties as in \p{eta2}
\be\label{QQQ}
\delta \eta = \epsilon+\epsilon \eta \partial_t \eta, \qquad
\delta \lambda = \epsilon \eta \partial_t \lambda.
\ee
The superfields $\xi(t,\theta),\bar\xi(t,\theta)$ as well as the genuine superfields $\psi(t,\theta),\nu(t,\theta)$ can be expressed in terms of the Goldstone
superfield $\eta(t,\theta)$ and matter fermionic superfield $\lambda(t,\theta)$:
\be\label{inverse}
\xi = \eta - \im \lambda -\im \eta\lambda \big( \dot \eta -\im \dot \lambda\big), \qquad
\bar\xi = \eta + \im \lambda +\im \eta\lambda \big( \dot \eta +\im \dot \lambda\big),
\ee
\bea\label{fin2}
&&
\psi =\frac{\eta - \im \lambda}{1+\big( D\eta- \im D\lambda\big)^2}- \im \eta\lambda \big(\dot \eta- \im \dot \lambda\big)
\frac{1- \big( D\eta- \im D\lambda\big)^2}{\Big[ 1+\big( D\eta- \im D\lambda\big)^2\Big]^2},\quad
\bar\psi = \big(\psi\big)^\dagger, \\
&&
\nu =\big( D\eta- \im D\lambda\big)\bigg( \frac{\eta- \im \lambda}{1+\big( D\eta- \im D\lambda\big)^2}- \im
\frac{2\eta\lambda \big(\dot \eta- \im \dot \lambda\big)}{\Big[ 1+\big( D\eta- \im D\lambda\big)^2\Big]^2}\bigg), \quad
\bar\nu = \big(\nu\big)^\dagger. \nn
\eea

\subsection{Relations between the two cases}
Since the relations between the pairs of superfields $(\xi_1,\xi_2)$ and $(\eta, \lambda)$ as well as between
$(\xi,\bxi)$  and $(\eta, \lambda)$ are invertible, one can express
the spinor fields $(\xi_1,\xi_2)$ through $(\xi,\bxi)$ and vice versa
\bea\label{twocases1}
&&
\xi= \frac{1}{2}\left(1-\im\right) \xi_1+ \frac{1}{2}\left( 1+\im\right) \xi_2+ \frac{1}{2}\xi_1 \xi_2 \left(
\dot{\xi}_1 - \dot{\xi}_2\right), \\
&&
\bar\xi= \frac{1}{2}\left(1+\im\right) \xi_1+ \frac{1}{2}\left( 1-\im\right) \xi_2+ \frac{1}{2}\xi_1 \xi_2 \left(
\dot{\xi}_1 - \dot{\xi}_2\right), \nn
\eea
\bea\label{twocases2}
&&
\xi_1= \frac{1}{2}\left(1+\im\right) \xi + \frac{1}{2}\left( 1-\im\right) \bar\xi + \frac{1}{2}\xi \bar\xi \left(
\dot{\xi} - \dot{\bar\xi}\right), \\
&&
\xi_2= \frac{1}{2}\left(1-\im\right) \xi + \frac{1}{2}\left( 1+\im\right) \bar\xi + \frac{1}{2}\xi \bar\xi \left(
\dot{\xi} - \dot{\bar\xi}\right).\nn
\eea
Thus, at least in one dimension, the two cases are completely equivalent, because they are related by invertible
fields redefinitions \p{twocases1}, \p{twocases2}.
The only difference between these cases lies in the diverse definition
of the invariant actions \p{action-tilde1}, \p{action-tilde}.
In fact, these two actions are not the most general ones. We will demonstrate this by constructing the most general action for the Goldstone fermion
$\eta(t,\theta)$ and the matter fermionic superfield $\lambda(t,\theta)$.

\setcounter{equation}{0}
\section{Nonlinear realizations approach}
In this section we provide an alternative description of the system with a partial breaking
of the global $N=2,d=1$ supersymmetry within the nonlinear realizations approach.
It turns out that in the present case this approach is more suitable for the construction of
the most general superfield action.

\subsection{NLR approach in one dimension}
We start with the $N=2, d=1$ Poincar\'{e} superalgebra with one central charge generator $Z$
\be\label{algebra}
\big \{ Q,Q \big\} = 2P, \quad \big \{ S,S \big\} = 2P, \quad \big \{ Q,S \big\} = 2Z.
\ee
Introducing a coset element $g$ as
\be\label{coset}
g = e^{tP} e^{\theta Q} e^{q Z} e^{\eta S},
\ee
and calculating the expression $g^{-1} dg = \omega_P P + \omega_Q Q + \omega_Z Z + \omega_S S$,
one finds the explicit structure for the Cartan forms
\be\label{Cartan}
\omega_P = dt - d \theta \theta - d \eta \eta, \quad  \omega_Q = d \theta, \quad
\omega_Z = d q - 2 d \theta \eta, \quad \omega_S = d \eta.
\ee
The covariant derivatives can be found in a standard way, and now they read
\be\label{der}
\nabla_{\theta} = D + \eta D \eta \nabla_t, \quad
\nabla_t = E^{-1} \partial_t,
\ee
where the explicit forms of the einbein $E$ and its inverse $E^{-1}$ are
\be\label{DE}
E = 1+ \eta \partial_t \eta, \quad E^{-1} = 1 - \eta \nabla_t \eta.
\ee
These derivatives satisfy the following (anti)commutation relations
\bea\label{alg-der}
&&
\big \{ \nabla_{\theta}, \nabla_{\theta} \big \} = 2 \big( 1 +  \nabla_{\theta} \eta \nabla_{\theta} \eta\big) \nabla_t, \\
&&
\big [ \nabla_t, \nabla_{\theta} \big ] = 2 \nabla_t \eta \nabla_{\theta} \eta \nabla_t. \nn
\eea
Acting on the coset element $g$ \p{coset} from the left by different elements $g_0$ of
the $N=2, d=1$ Poincar\'{e} supergroup, one can find the transformation properties of the coordinates $(t, \theta)$
and the Goldstone superfield $\eta(t, \theta)$.
In particular, under both unbroken $Q$- and broken $S$-supersymmetries one gets:
\begin{itemize}
\item Unbroken supersymmetry ($g_0=e^{\epsilon Q}$\,):
\be\label{susyQ}
\delta_Q t = - \epsilon \theta, \quad
\delta_Q \theta=\epsilon,
\ee
\item Broken supersymmetry ($g_0=e^{\eps S}$\,):
\be\label{susyS}
\delta_S t = - \eps \eta, \quad \delta_S q = -2 \eps \theta, \quad
\delta_S \eta = \eps.
\ee
\end{itemize}
The final step is to express the Goldstone superfield $\eta(t, \theta)$ in terms of the bosonic superfield $q(t, \theta)$
by imposing the constraint \cite{ih}
\be\label{ih}
\omega_Z|_{d\theta} =0 \qquad \Rightarrow \qquad \eta=\frac{1}{2}\nabla_{\theta} q.
\ee
In order to describe the matter spinor superfield $\lambda(t, \theta)$ within the
nonlinear realizations approach, it is necessary to postulate  its transformation properties under the $S$-supersymmetry
\be\label{sl}
\delta_S \lambda =0.
\ee
It is obvious that, due to the transformation of the coordinate $t$ under the $S$-supersymmetry
in \p{susyS}, the variation of $\lambda(t, \theta)$ reads
\be\label{del-lam}
\delta \lambda = \eps \eta \partial_t \lambda.
\ee
One may also define the bosonic superfield $\phi(t, \theta)$ with the transformation property
\be\label{sphi}
\delta_S \phi =0,
\ee
which is related with $\lambda(t, \theta)$ in the same way as in \p{ih}
\be\label{ih2}
\lambda = \frac{1}{2} \nabla_{\theta} \phi.
\ee

\subsection{General action}
The most general {\it Ansatz} for the $N=1$ superfield action describing the Goldstone fermionic superfield $\eta(t,\theta)$
and the matter fermionic superfield $\lambda(t,\theta)$ and having no dimensional constants reads
\be\label{S1}
S= \int dt d\theta \big( 1+ \eta \dot \eta\big) \Big[ \eta\; F_1^{cov} + \lambda\; F_2^{cov} + \eta \dot \eta \lambda\; F_3^{cov} +
\eta \lambda \nabla_t \lambda\; F_4^{cov} \Big],
\ee
where $F_1^{cov}, F_2^{cov}, F_3^{cov}, F_4^{cov}$ are arbitrary functions depending on $\nabla_{\theta}\eta$
and $\nabla_{\theta}\lambda$ only.
The reason for such a form of {\it Ansatz} is quite understandable:
\begin{itemize}
\item with respect to $S$-supersymmetry transformation, $\delta_S t =- \eps \eta,\; \delta_S \eta = \eps,\; \delta_S \lambda=0$,
the ``improved'' measure $dt d\theta \big( 1+ \eta \dot \eta\big)$ is invariant;
\item the functions $F_1^{cov}, F_2^{cov}, F_3^{cov}, F_4^{cov}$ as well as $\nabla_t \lambda$ are also invariant with respect
to the broken $S$-supersymmetry;
\item the functions $F_1^{cov}, F_2^{cov}, F_3^{cov}, F_4^{cov}$ have to be further restricted by imposing the invariance
of the action \p{S1} with respect to the broken $S$-supersymmetry.
\end{itemize}
The variation of the action \p{S1} with respect to $S$-supersymmetry reads
\be\label{S2}
\delta_S S =  \int dt d\theta \;\eps \Big[ \big( 1 + \eta \dot \eta\big) F_1^{cov}
+ \dot \eta \lambda F_3^{cov} + \lambda \dot \lambda F_4^{cov} \Big].
\ee
Thus, the function $F_2^{cov}$ may be chosen to be an arbitrary function of its arguments.

\noindent
To fix the functions $F_1^{cov}, F_3^{cov}, F_4^{cov}$
from the condition $\delta_S S=0$ one has, firstly,
to integrate over $\theta$ in \p{S2} and then to replace the arguments of these functions,
$\nabla_\theta \eta$ and $\nabla_\theta \lambda$, by
\be\label{ss1}
\nabla_\theta \eta = \big( 1+\eta \dot \eta\big) D\eta \equiv \big( 1+\eta \dot \eta\big) x, \quad
\nabla_\theta \lambda = D\lambda +\eta \dot \lambda D\eta \equiv y + \eta \dot \lambda x.
\ee
Hence, the explicit expression for each of $F^{cov}[\nabla_\theta \eta, \nabla_\theta \lambda ]$ takes the form
\be\label{ss2}
F^{cov}[\nabla_\theta \eta, \nabla_\theta \lambda ] = F[x,y] + \eta\dot \eta x (F[x,y])_x + \eta \dot \lambda x (F[x,y])_y,
\ee
where, by definition,
\be\label{ss3}
x \equiv D\eta, \qquad y \equiv D\lambda.
\ee
Now
\begin{itemize}
\item performing the integration in \p{S2} over $\theta$, \\
\item expanding the functions $F_1^{cov}, F_3^{cov}, F_4^{cov}$ as in \p{ss2}, \\
\item collecting the terms which are linear and cubic in the fermions, \\
 \item and assuming that the right hand side of variation is a total derivative
with respect to $t$ of an arbitrary function depending on $x,y$,
\end{itemize}
we will get the following three equations
\bea\label{eq4}
\Big[ \big(1+x^2\big) F_1 \Big]_x - y F_3 =a, && \qquad (a)  \nn\\
y \Big[ \big( 1+x^2\big) F_1 \Big]_{yy} +\big( y^2 F_4 \big)_y =0, && \qquad (b) \\
\big(F_3\big)_y +\big(F_4\big)_x=0, && \qquad (c) \nn
\eea
where $a$ is an arbitrary constant. The equation $(\ref{eq4}b)$ may be integrated once, giving
\be\label{eq5}
y \big( \widetilde F_1\big)_y - \widetilde F_1 +y^2 F_4 = G[x],
\ee
where
\be\label{eq6}
\widetilde F_1 \equiv \big( 1+ x^2\big) F_1,
\ee
and $G[x]$ is an arbitrary function depending on $x$. Then the  equation $(\ref{eq4}a)$ reads
\be\label{eq7}
\big( \widetilde F_1\big)_x - y F_3 =a.
\ee
Differentiating the equation \p{eq5} over $x$ and the equation \p{eq7} over $y$, and using the equation  $(\ref{eq4}c)$
one may get
\be\label{eq8}
\big(G[x]\big)_x = -a \quad \Rightarrow \quad G[x]= -a x -b,
\ee
where $b$ is a new constant.
Finally, representing the function $\widetilde F_1$ as
\be\label{eq9}
\widetilde F_1 = a x + b + y \widehat F_1
\ee
we will finish with the equations
\be\label{eq10}
\big( \widehat F_1\big)_y+ F_4=0, \qquad \big(\widehat F_1\big)_x - F_3=0,
\ee
which define the functions $F_3, F_4$ in terms of an arbitrary function $\widehat F_1[x,y]$.

\noindent
Thus, the general action \p{S1} is invariant under the broken $S$-supersymmetry,
if the functions $F_1, F_3$ and $F_4$ are expressed through arbitrary function $F_1$ as
\be\label{eq11}
F_1 = \frac{ax + b}{1+x^2} +\frac{y}{1+x^2} \widehat F_1,\quad F_3 = \big(\widehat F_1\big)_x,\quad
F_4 = -\big( \widehat F_1\big)_y.
\ee
Thus, the most general action for the present system is defined up to two arbitrary functions $\widehat F_1$ and $F_2$.

\noindent
Let us note that the constant $b$ can be chosen equal to zero, because the action corresponding
to the Lagrangian
\be
L = \frac{b}{1+x^2}
\ee
is trivial.
Thus, the final result for the covariant functions $F^{cov}$ which enter the action \p{S1} reads
\be\label{fineq}
F_1^{cov} = a \frac{\nabla_\theta \eta}{1+\big(\nabla_\theta \eta\big)^2} +\frac{\nabla_\theta \lambda}{1+\big(\nabla_\theta \eta\big)^2} \widehat F_1^{cov},\quad
F_3^{cov} = \big(\widehat F_1\big)_{\nabla_\theta \eta},\quad
F_4^{cov} = -\big( \widehat F_1\big)_{\nabla_\theta \lambda},
\ee
with $\widehat F_1^{cov}$ and $F_2^{cov}$ being arbitrary
functions on $\nabla_\theta \eta$ and $\nabla_\theta \lambda$
\be\label{arF}
\widehat F_1 = \widehat F_1 [\nabla_\theta \eta, \nabla_\theta \lambda ], \qquad F_2 = F_2 [\nabla_\theta \eta, \nabla_\theta \lambda].
\ee

\setcounter{equation}{0}
\section{Interesting cases}
Having at hands the most general expression for the action \p{S1}, \p{fineq} invariant with respect to both $Q$- and $S$-supersymmetries,
it is interesting to visualize the functions $F_1, F_2$ which reproduce the {\it FPS} actions \p{action-tilde1}, \p{action-tilde}.
Comparing \p{fin1} and \p{fin2} one may conclude, that these two {\it FPS} cases are related through the
substitutions: $\lambda \rightarrow - \im \lambda, y \rightarrow -\im y$ (where $y=D \lambda$).
Therefore, it is enough to consider the first case, with $d_{111}=d_{222}=1$, only.

In order to simplify everything, it is useful to represent the integrand in \p{S1} as follows
\be\label{res1}
L= \eta F_1+ \lambda F_2 +\lambda \eta \dot \eta \big( x F_2 +\widehat F_1\big)_x-
\eta\lambda \dot \lambda \big( x F_2 + \widehat F_1 \big)_y,
\ee
with
\be
F_1 \equiv a \frac{x}{1+x^2}+\frac{y}{1+x^2}\widehat F_1,
\ee
where the functions $\widehat F_1 $ and $F_2$ depend on $x,y$ variables \p{ss3}.
Comparing the integrands in the actions \p{action-tilde1} and \p{res1}, we will get for the two special choices of the constant parameters
$e_a$ and $C_{ab}$ of the {\it FPS} action the following result
\bea
(a) && if \quad e_2=C_{ab}=0,\qquad then \quad L_{FPS} = \nu_1 \quad \Rightarrow  \nn\\
&&
 F_1=F_2= \frac{x+y}{1+(x+y)^2},\;\; a=1 \label{ac1}\\
(b) && if \quad e_1=e_2=C_{ab}=1,\qquad then \quad L_{FPS} =\psi_1 D\psi_2 + \nu_1 D \nu_2 \quad \Rightarrow \nn\\
&&
F_1=F_2= \frac{(x-y)(1+x^2-y^2)}{(1+(x-y)^2)(1+(x+y)^2)}, \;\; a=1. \label{ac2}
\eea
Unfortunately, the explicit form of the functions $F_1, F_2$ which corresponds to {\it FPS} actions \p{ac1}, \p{ac2}
is not informative enough to understand why these actions were selected.

Let us remind that the general action
\be\label{S11}
S= \int dt d\theta \big( 1+ \eta \dot \eta\big) \Big[ \eta\; F_1^{cov} + \lambda\; F_2^{cov} + \eta \dot \eta \lambda\; F_3^{cov} +
\eta \lambda \nabla_t \lambda\; F_4^{cov} \Big],
\ee
with the restrictions
\be\label{fineqqq}
F_1^{cov} = a \frac{\nabla_\theta \eta}{1+\big(\nabla_\theta \eta\big)^2} +\frac{\nabla_\theta \lambda}{1+\big(\nabla_\theta \eta\big)^2} \widehat F_1^{cov},\quad
F_3^{cov} = \big(\widehat F_1\big)_{\nabla_\theta \eta},\quad
F_4^{cov} = -\big( \widehat F_1\big)_{\nabla_\theta \lambda}
\ee
contains two arbitrary functions $F_1^{cov}$ and $F_2^{cov}$.
So, as we expected, the invariance with respect to additional, spontaneously broken $N=1$ supersymmetry does not fix the action
in the many-particles case, in contrast with the one-particle case.

\section{Conclusion}
In this work we analysed in details the application of the {\it FPS} approach in $d=1$.

We demonstrated that the resulting system describes the $N=2, d=1$ supersymmetric action for
two particles in which one of $N=1$ supersymmetries is spontaneously broken.
The final actions possess the same features as the {\it FPS} ones.

Using the nonlinear realization approach we reconsider the system in the basis
where only one superfield has the Goldstone nature while the second superfield
can be treated as the matter one.
Having at hands the transformations relating the two selected {\it FPS} cases with
our more generic one, we established the field redefinitions which relate these cases.
Namely, in this basis, the two {\it FPS} supermultiplets are related by the redefinition of the matter superfield
$\lambda \rightarrow -\im \lambda$.
This analysis leads to the conclusion that the only difference between two {\it FPS} cases lies in a various choice of actions,
while the supermultiplets specified by the {\it FPS} constraints are really the same.

Going further on with the nonlinear
realization approach, we constructed the most general action for the system of two $N=1$
superfields possessing one additional hidden spontaneously broken $N=1$ supersymmetry.
This action contains two arbitrary functions and reduces to the {\it FPS} actions upon specification of these functions.

Unfortunately, the exact form of these functions corresponding to {\it FPS} actions
is not very informative and it gives no hint about the reason why the {\it FPS} cases were selected.
Of course, our consideration was strictly one-dimensional and therefore
we cannot argue that all features we discussed will appear in the generalized
supersymmetric Born-Infeld theory constructed in \cite{FPS,FPS1}:

Nevertheless, we believe that our results and the generality of the nonlinear
realization approach are yield reasonable tools to reconsider the questions:
\begin{itemize}
 \item whether the supermultiplets with different constants $d_{abc}$ are really different?
 \item which additional properties select the {\it FPS} actions?
 \end{itemize}
in four dimensions.

\section*{Acknowledgments}
The work of S.K. was supported by RSCF grant 14-11-00598.
The work of A.S was partially supported by RFBR grants 15-02-06670 and 15-52-05022 Arm-a.

\end{document}